\documentclass{edbk} 
  \usepackage{edbkps}

\normallatexbib

\usepackage[dvips]{graphicx}

\newcommand{\ket} [1] {\vert #1 \rangle}
\newcommand{\bra} [1] {\langle #1 \vert}

\newcommand{\proj}[1]{\ket{#1}\bra{#1}}
\newcommand{\mean}[1]{\langle #1 \rangle}                                       

\begin{document}

\articletitle[Continuous-variable cloning]{Quantum cloning\\ 
with continuous variables$^*$\thanks{To appear in {\it Quantum Information 
with Continuous Variables}, edited by S. L.Braunstein and A. K. Pati,
(Kluwer Academic, 2002).}}

\author{Nicolas J. Cerf}
\affil{Ecole Polytechnique, CP 165, Universit\'e Libre de Bruxelles,
1050 Brussels, Belgium}
\email{ncerf@ulb.ac.be}

\section[]{Introduction}

Quantum information theory has developed dramatically over the past
de\-cade, driven by the prospects of quantum-enhanced
communication and computation systems. Among the most striking
successes, one finds for example the discovery of quantum
factoring, quantum key distribution, or quantum teleportation. 
Most of these concepts were initially developed for
discrete quantum variables, in particular quantum bits, which have
now become the symbol of quantum information. Recently,
however, a lot of attention has been devoted to investigating the
use of \emph{continuous-variable} systems in quantum informational or
computational processes. Continuous-spectrum quantum variables, 
for example the quadrature components of a light mode, may be easier 
to manipulate than quantum bits. It is actually sufficient 
to process squeezed states of light into linear optics circuits 
in order to perform various quantum information processes 
over continuous variables \cite{brau98:qerr}. As reported in the present book,
variables with a continuous spectrum have been shown to be useful 
to carry out quantum teleportation, quantum entanglement purification,
quantum error correction, or even quantum computation. 

In this Chapter, the issue of \emph{cloning} a continuous-variable 
quantum system will be analyzed, and a Gaussian cloning transformation 
will be introduced.
Cloning machines, that is, transformations that achieve the best 
approximate copying of a quantum state compatible with the
no-cloning theorem, have been a fundamental research topic 
over the last five years (see e.g. \cite{brau01} for an overview).
This question is of particular significance given the close connection
between quantum cloning and quantum cryptography: using an optimal
cloner generally makes it possible to obtain a tight bound on the
best individual eavesdropping strategy in a quantum cryptosystem.
This provides a strong incentive to investigating continuous-variable
cloning in view of the recent proposals for quantum key distribution 
relying on continuous (Gaussian) key carriers
\cite{cerf00_qdgk,gros02_coherent}.

Here, we will focus on a Gaussian cloning transformation,
which copies equally well any two canonically conjugate continuous variables 
such as the two quadrature components of a light mode \cite{cerf00:cont}. 
More precisely, it achieves the {\em optimal} cloning 
of a continuous variable that satisfies the requirement 
of covariance with respect to displacements and rotations in phase space.
Consequently, this cloner duplicates all coherent states with a same
fidelity ($F=2/3$). The optical implementation of this cloner and its
extension to $N$-to-$M$ cloners will also be discussed. Finally,
the use of this cloner for the security assessment
of continuous-variable quantum key distribution schemes will be sketched.

\section[]{Limits on optimal cloning}

Let us start by stating the problem of continuous-variable cloning 
in physical terms. 
Consider, as an example of canonically conjugate continuous variables,
the quadrature components of a light mode, denoted as $x$ and $p$.
This notation reflects the fact that $x$ and $p$ behave just like the
position and momentum of a particle in a one-dimensional space,
namely their commutator is $[x,p]=i$ (we put $\hbar=1$ in this paper).
If the wave function is a Dirac delta
function---the particle is fully localized in {\em position} space, 
then $x$ can be measured exactly, and several perfect copies of the
system can be prepared. However, such a cloning process fails
to exactly copy non-localized states, {\it e.g.},  momentum
states. Conversely, if the wave function is a plane wave with
momentum $p$---the particle is localized in {\em momentum} space, then
$p$ can be measured exactly and one can again prepare several perfect copies
of this plane wave. However, such a ``plane-wave cloner'' is then
unable to copy position states exactly. In short, it is impossible
to copy perfectly the eigenstates of two conjugate variables 
such as $x$ and $p$: this is essentially the content 
of the so-called {\em no-cloning} theorem \cite{woot82,diek82}.

In the next Section, we will show that a {\em cloning} transformation
can nevertheless be found that provides two copies of a continuous
system, but at the price of a non-unity cloning fidelity. In other
words, the cloning machine yields two {\em imperfect} copies of the system.
Before describing this cloning machine in details, let us find
a lower bound on the cloning-induced noise by exploiting a connection
with measurement theory. More specifically, we make use of the fact
that measuring $x$ on one clone and $p$ on the other clone
cannot beat the optimal joint measurement of $x$ and $p$ on the
original system \cite{cerf00:coherent}. 
It is known that such a joint measurement of a pair of conjugate observables
on a single quantum system
obeys an inequality akin to the Heisenberg uncertainty relation but with
an extra contribution to the minimum variance \cite{arth65}.
Denoting by $x$ and $p$ the two quadratures of the input mode, and by
$X$ and $P$ the corresponding jointly measured output quadratures, we have
\begin{eqnarray}
X=x+n_x \mathletter{a}\\
P=p+n_p \mathletter{b}
\end{eqnarray}
where $n_x$ and $n_p$ stand for the excess noise that we have 
on the measured quadratures. Since we consider a joint measurement,
the variables $X$ and $P$ must commute: they can be viewed respectively
as the $x$ and $p$ quadratures of two distinct modes. Thus, we have
\begin{equation}
[X,P]=[x,p]+[x,n_p]+[n_x,p]+[n_x,n_p]=0
\end{equation}
Assuming that the excess noises $n_x$ and $n_p$ are independent 
of the input quadratures, {\it i.e.}, $[x,n_p]=[n_x,p]=0$, 
we get $[n_x,n_p]=-i$, implying that $n_x$ and $n_p$
must obey an uncertainty relation.
Specifically, any attempt to measure $x$ and $p$ simultaneously 
on a quantum system is constrained by the inequality
\begin{equation} \label{addednoise}
\Delta n_x \; \Delta n_p \geq 1/2
\end{equation}
where $\Delta n_x^2$ and $\Delta n_p^2$ denote the variances of
the excess noises originating from the joint measurement device.
If the variances of the $x$ and $p$ quadratures of the input state 
are denoted by $\delta x^2$ and $\delta p^2$, respectively,
we thus have for the variances of the measured values
$\Delta X ^2 = \delta x^2 + \Delta n_x^2$ and
$\Delta P ^2 = \delta p^2 + \Delta n_p^2$.
As a consequence, the Heisenberg uncertainty relation 
$\delta x\; \delta p \geq 1/2$
together with inequality (\ref{addednoise}) implies the relation \cite{arth65}
\begin{equation}
\Delta X \; \Delta P \geq 1
\end{equation}
where we have used the inequality $a^2+b^2\geq 2\sqrt{a^2 b^2}$.
Thus, the best possible joint measurement of $x$ and $p$
with a same precision on both quadratures
of a coherent state ($\delta x^2=\delta p^2=1/2$)
gives 
\begin{equation}
\Delta X^2= \Delta P^2=1 
\end{equation}
Compared with the vacuum noise, we note that the joint measurement
of $x$ and $p$ effects an additional noise of minimum variance 1/2,
so that the measured values suffer {\em twice} the vacuum noise.

Inequality (\ref{addednoise}) immediately translates into
a lower bound on the cloning-induced noise variance \cite{cerf00:coherent}. 
If we assume
that the device that is used in order to perform the joint measurement
of $x$ and $p$ is actually a cloning machine followed by two
measuring apparatuses ($x$ being measured on one clone and $p$
on the other clone), we conclude that the variance of the noise added 
by this cloning machine cannot be lower than 1/2 in order to comply
with Eq. (\ref{addednoise}), that is
\begin{equation}
\Delta n_x^2 = \Delta n_p^2 \geq 1/2
\end{equation}
(We require here the same noise level on $x$ and $p$.)
This can also be shown explicitly by
writing the canonical transformation of the cloner \cite{gros01}. 
Denoting by $X_{a(b)}$ and $P_{a(b)}$ the two quadratures of
the output mode $a$ (resp. $b$), we have
\begin{eqnarray}
X_a=x+n_{x,a} \mathletter{a}\\
P_a=p+n_{p,a} \mathletter{b}\\
X_b=x+n_{x,b} \mathletter{c}\\
P_b=p+n_{p,b} \mathletter{d}
\end{eqnarray}
where $x$ and $p$ are the two quadratures of the input mode
and $n_{x/p,a/b}$ stand for the excess noises. Since the clones
are carried by different modes ($a$ and $b$), we have
$[X_a,P_b]=[X_b,P_a]=0$. Assuming, as before, that the 
excess noises are independent of the input mode, we get
$[n_{x,a},n_{p,b}]=[n_{x,b},n_{p,a}]=-i$. This gives rise
to two no-cloning uncertainty relations 
\begin{eqnarray}  \label{noclonineqa}
\Delta n_{x,a} \; \Delta n_{p,b} \geq 1/2  \mathletter{a}\\
\Delta n_{x,b} \; \Delta n_{p,a} \geq 1/2  \mathletter{b} \label{noclonineqb}
\end{eqnarray}
which constrain the excess noise variances
$\Delta n_{x/p,a/b}^2$ of the two clones \cite{cerf00:cont,gros01}. 
Consequently, if the cloning process
induces a small position (momentum) error on the first copy,
then the second copy is necessarily affected by a large
momentum (position) error. The Gaussian cloner we will discuss
in the next Session saturates these inequalities and is symmetric
in $a$ and $b$ (and in $x$ and $p$):
\begin{equation}  \label{saturate}
\Delta n_{x,a}^2=\Delta n_{p,a}^2=\Delta n_{x,b}^2=\Delta n_{p,b}^2=1/2
\end{equation}
To simplify the notation, we will denote this cloning-induced excess noise
variance as $\sigma^2$ in the following.

\section[]{Gaussian cloning transformation}

We will define a class of cloning machines 
that yield two imperfect copies of a continuous-variable system, 
the underlying cloning transformation being {\em covariant} with respect to
displacements in phase space $(x,p)$. 
By this, we mean that any two input states that are related by a displacement
result in copies that are related in the same way; 
hence, the resulting cloning fidelity
is invariant under displacements in phase space.
Specifically, let us seek for a displacement-covariant transformation which
duplicates with a same fidelity all coherent states $\ket{\psi}$.
Thus, if two input states are identical up to a
displacement $\hat D(x',p')=e^{-ix' \hat p} e^{ip' \hat x}$, then their
respective copies should be identical up to the same displacement. 
Denoting by $\mathcal{H}$ the Hilbert
space corresponding to a single system, cloning can be defined as a
completely-positive trace-preserving linear map ${\mathcal{C}}:
\mathcal{H} \to \mathcal{H}^{\otimes 2}: \proj{\psi} \to {\mathcal{C}}
(\proj{\psi})$ such that
\begin{equation}
\label{eq:cov}
{\mathcal{C}} \left[ \hat{D}(x',p')\proj{\psi} \hat{D}^{\dagger}(x',p')
 \right] \\
= \hat{D}(x',p')^{\otimes 2} \; {\mathcal{C}} (\proj{\psi}) \; 
  \hat{D}^{\dagger}(x',p')^{\otimes 2}
\end{equation}
for all displacements $\hat D(x',p')$.

\begin{figure}
\begin{center}
\includegraphics[width=3.0in,angle=0]{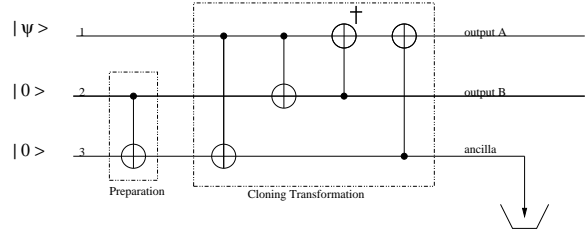}
\caption{Quantum circuit for the continuous-variable cloning transformation.
It consists of four {\sc c-not} gates preceeded by a preparation stage. 
Here, the ancillae are prepared in the state given by Eq. (\ref{eq_vacuum}).
See \cite{cerf00:cont,cerf01:capri}.}
\label{fig:ccircuit}
\end{center}
\end{figure}

As shown in \cite{cerf00:cont}, this cloning map can be achieved
via a unitary transformation ${\hat{\cal U}}$ acting on three modes:
the input mode (variable 1) supplemented with two auxiliary
modes, the blank copy (variable 2) and an ancilla (variable 3).
The two auxiliary variables must be initially prepared in the joint state
\begin{equation}   \label{eq_chi}
|\chi\rangle_{2,3} =
\int\!\!\int_{-\infty}^{\infty} dx \, dp \; f(x,p) \;
|\Psi(x,-p)\rangle_{2,3}
\end{equation}
where $f(x,p)$ is an (arbitrary) complex amplitude function, and
\begin{equation}   \label{eq_EPR}
|\Psi(x,p)\rangle = {1\over\sqrt{2\pi}} \int_{-\infty}^{\infty}
dx'\; {\rm e}^{ipx'} \; |x'\rangle |x'+x\rangle
\end{equation}
are the EPR states (the maximally-entangled states 
of two continuous variables).
The cloning transformation is defined as
\begin{equation}   \label{eq_defU}
{\hat{\cal U}}_{1,2,3} = {\rm e}^{-i({\hat x_3}-{\hat x_2}){\hat p_1} } \;
{\rm e}^{-i{\hat x_1}({\hat p_2}+{\hat p_3}) }
\end{equation}
where ${\hat x_k}$ (${\hat p_k}$) is the position (momentum) operator
for variable $k$. As shown in Fig.~\ref{fig:ccircuit},
this can be interpreted as a sequence of four 
continuous-variable controlled-{\sc not} ({\sc c-not}) gates,
each being defined as the unitary transformation
${\rm e}^{-i{\hat x_k} {\hat p_l}}$ with $k$ ($l$) referring to the
control (target) variable~\cite{brau98:qerrprl}.

Remarkably, Eq.~(\ref{eq_defU})
coincides with the discrete {\sc c-not} gate sequence that achieves
the qubit cloning transformation~\cite{buze96}, up to a sign ambiguity
originating from the fact that a continuous {\sc c-not} gate
is not equal to its inverse.  After applying ${\hat{\cal U}}$
to the state $|\psi\rangle_1 |\chi\rangle_{2,3}$, we get the joint state
\begin{equation}   \label{eq_phi}
\int\!\!\int_{-\infty}^{\infty} dx \, dp \; f(x,p) \;
\hat D(x,p)\ket{\psi}_1 \; |\Psi(x,-p)\rangle_{2,3}
\end{equation}
where variables 1 and 2 are taken as the two outputs of the cloner
(clones $a$ and $b$), while variable 3 (the ancilla) must simply be
traced over. This is a peculiar state in that it can be reexpressed
in a similar form by exchanging the two clones, namely
\begin{equation}   \label{eq_phi2}
\int\!\!\int_{-\infty}^{\infty} dx \, dp \; g(x,p) \;
\hat D(x,p)\ket{\psi}_2 \; |\Psi(x,-p)\rangle_{1,3}
\end{equation}
with 
\begin{equation}    \label{eq_dft}
g(x,p)= {1\over 2\pi}
\int\!\!\int_{-\infty}^{\infty} dx' \, dp' \;
{\rm e}^{i(px'-xp')} \; f(x',p')
\end{equation}
being the two-dimensional Fourier transform of $f(x,p)$.
The resulting state of the individual clones can then be written as
\begin{eqnarray}
\rho_a=
\int\!\!\int_{-\infty}^{\infty} dx \, dp \; |f(x,p)|^2 \;
\hat D(x,p)\ket{\psi}\bra{\psi} \hat D^{\dagger}(x,p) 
\mathletter{a} \\
\rho_b=
\int\!\!\int_{-\infty}^{\infty} dx \, dp \; |g(x,p)|^2 \;
\hat D(x,p)\ket{\psi}\bra{\psi} \hat D^{\dagger}(x,p) 
\mathletter{b}
\end{eqnarray}
which is consistent with tracing Eq. (\ref{eq:cov}) over any one of the clones.
Thus, the clones are affected by position and momentum errors that are
distributed according to $|f(x,p)|^2$ and $|g(x,p)|^2$.
A central point here is that 
interchanging the two clones amounts to substitute the function
$f$ with its two-dimensional Fourier transform $g$. This property is crucial
as it ensures that the two copies suffer from {\em complementary}
position and momentum errors.
Indeed, one can check \cite{cerf00:cont} that the four excess noise variances
defined as 
\begin{eqnarray}
&&\Delta n_{x,a}^2 =\int\!\!\int_{-\infty}^{\infty} dx \, dp \; x^2 \, |f(x,p)|^2, \mathletter{a}\\ 
&&\Delta n_{p,a}^2 =\int\!\!\int_{-\infty}^{\infty} dx \, dp \; p^2 \, |f(x,p)|^2   \mathletter{b}\\
&&\Delta n_{x,b}^2 =\int\!\!\int_{-\infty}^{\infty} dx \, dp \; x^2 \, |g(x,p)|^2, \mathletter{c}\\ 
&&\Delta n_{p,b}^2 =\int\!\!\int_{-\infty}^{\infty} dx \, dp \; p^2 \, |g(x,p)|^2   \mathletter{d}
\end{eqnarray}
obey the no-cloning inequalities (\ref{noclonineqa})
and (\ref{noclonineqb}). (Here, we assume that the first-order moments
of $|f(x,p)|^2$ and $|g(x,p)|^2$ vanish, that is, the clones
are not biased.)

Within this class of cloning machines parametrized by $f(x,p)$,
a particularly simple {\em rotation-covariant} cloner
can be found that provides two {\em identical} copies of a continuous system
with the {\em same} error distribution in position and momentum.
It corresponds to the choice $f(x,p) = g(x,p) 
= {\rm e}^{- (x^2+p^2)/2 }/\sqrt{\pi}$.
This cloner is named ``Gaussian'' as it effects Gaussian-distributed
position- and momentum-errors on the input mode: the excess noise
on both clones is distributed as
${\rm e}^{- (x^2+p^2) }/\pi$, that is,
as a bi-variate rotational-invariant Gaussian of variance $\sigma^2=1/2$.
This cloner is {\em optimal}, as it satisfies Eq. (\ref{saturate}).
Here, the two auxiliary variables must be prepared in the state
\begin{equation}  \label{eq_vacuum}
|\chi\rangle_{2,3} = {1\over\sqrt{\pi}}
\int\!\!\int_{-\infty}^{\infty} dy\, dz \; {\rm e}^{-{y^2+z^2\over 2}}
\; |y\rangle_2  \; |y+z\rangle_3
\end{equation}
which is simply the product vacuum state $|0\rangle_2 |0\rangle_3$ 
processed by a {\sc c-not} gate ${\rm e}^{-i{\hat x_2}{\hat p_3}}$.
The resulting transformation effected by ${\hat {\cal U}}$
on an input position state $|x\rangle$ is thus given by
\begin{eqnarray}  \label{eq_U}
|x\rangle_1 |\chi\rangle_{2,3} \to {1\over\sqrt{\pi}}
\int\!\!\int_{-\infty}^{\infty} dy\, dz \; {\rm e}^{-{y^2+z^2\over 2}} 
|x+y\rangle_1  |x+z\rangle_2  |x+y+z\rangle_3
\end{eqnarray}
where the three variables denote the two clones and the ancilla,
respectively. For an arbitrary input state $|\psi\rangle$,
it is readily checked that this transformation outputs 
two clones whose individual states are Gaussian distributed
with a variance $\sigma^2=1/2$, namely
\begin{equation}
\label{eq:gaussmix}
\rho_a=\rho_b=\frac{1}{\pi}
\int\!\!\int_{-\infty}^{\infty} dx \; dp  \; e^{-(x^2+p^2)}
\hat D(x,p)\ket{\psi}\bra{\psi} \hat D^{\dagger}(x,p)  ,
\end{equation}
In particular, if the input is a coherent state $|\alpha\rangle$
with $\alpha=(x+ip)/\sqrt{2}$, it is easy to calculate the fidelity 
of this cloner by using $|\langle \alpha | \alpha' \rangle|^2
= \exp(-|\alpha-\alpha'|^2)$:
\begin{equation}
F=\langle\alpha|\rho_{a(b)}|\alpha\rangle
={1 \over1+\Delta n^2}={2\over 3}
\end{equation}
This cloning fidelity does not depend on $\alpha$, 
so this Gaussian cloner copies 
{\em all} coherent states with the same fidelity $2/3$.
It can be viewed as the continuous counterpart of the universal qubit
cloner \cite{buze96}, as its cloning fidelity
is invariant under rotations in phase space. 
The physical origin of the cloning noise becomes, however, much 
more evident in the case of continuous variables:
the Gaussian noise that affects the clones can simply be traced back 
to the Gaussian wave function 
of the two ancillary modes, see (\ref{eq_vacuum}).
This suggests that the noise that inevitably arises when cloning is
intrinsically linked to the vacuum fluctuations
of the auxiliary modes.    

Note finally that this formalism can easily be extended to the cloning of
squeezed states instead of coherent states \cite{cerf00:cont}. One simply
unsqueeze the state before cloning and then squeeze the clones again.
For any value of the squeezing parameter $r$, one can then define 
a Gaussian cloner that copies with fidelity 2/3 all squeezed states 
of which the same quadrature is squeezed by the same amount $r$. In contrast,
cloning these squeezed states using the rotation-covariant cloner defined above
results in a fidelity that decreases as $r$ increases.

\section[]{Optical implementation}

It is very instructive to write the cloning transformation
in the Heisenberg picture, that is, following the evolution of the
annihilation operators associated with the modes that are involved. 
Again, mode 1 denotes the input mode,
and modes 2 and 3 the ancillary modes. Mode 1' and 2' stand for
the two clones, while 3' is the ancilla that is traced over after cloning.
Here, $a_{j}=(x_{j}+i p_{j})/\sqrt{2}$ stands for the
annihilation operator for mode $j$.
We require that the cloning transformation conserves the mean
values, {\it i.e.}, $\mean{a_1'}=\mean{a_2'}=\mean{a_1}$, so
that the clones are centered on the original coherent state.
We also require that the cloning transformation is covariant under
rotations in phase space. It is shown in \cite{brau00}
that the optimal transformation satisfying these requirements is
\begin{eqnarray} \label{canonic}
a_1'&=&a_1+{a_2\over \sqrt{2}}+{a_3^{\dagger}\over \sqrt{2}}  \mathletter{a}\\
a_2'&=&a_1-{a_2\over \sqrt{2}}+{a_3^{\dagger}\over \sqrt{2}}  \mathletter{b}\\
a_3'&=&a_1^{\dagger}+\sqrt{2}\, a_3 \mathletter{c}
\end{eqnarray}
where mode 1 is initially prepared in an arbitrary coherent state
$\ket{\alpha}$, with $\alpha=(x+ip)/\sqrt{2}$,
while modes 2 and 3 are prepared in the vacuum state.
This transformation clearly satisfies the commutation rules
$[a_i',a_j']=\delta_{i,j}$
and yields the correct mean values $(x,p)$ for the two
clones (modes 1' and 2'). Also, one can easily check that the
quadrature variances of the clones are equal to twice the vacuum
noise, in accordance with the cloning excess noise variance
$\sigma^2=1/2$. This transformation actually coincides 
with the Gaussian cloner introduced
in the previous Section. Interestingly, we note here that
the state in which the ancilla $3$ is left after cloning
is centered on $(x,-p)$, that is
the {\em phase-conjugated} state $\ket{\alpha^*}$.
This means that, in analogy with the universal qubit cloner, 
the Gaussian cloner generates an ``anticlone'' (or time-reversed state)
together with the two clones.

\begin{figure}
\begin{center}
\includegraphics[width=3.0in,angle=0]{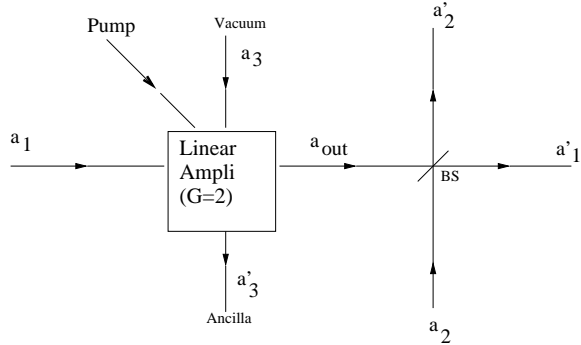}
\caption{Implementation of a Gaussian cloner using a
phase-insensitive linear amplifier and a $50{\rm:}50$ beam-splitter (BS).
See \cite{cerf01:capri}.}
\label{fig:ccircuit2}
\end{center}
\end{figure}

As suggested by the above transformation, 
a possible optical implementation of this Gaussian cloner 
consists in processing the input mode $a_1$ into a linear 
phase-insensitive amplifier \cite{cave82} of gain $G=2$:
\begin{equation}  \label{imp1}
a_{out}=\sqrt{2}\; a_{1}+ a_{3}^{\dagger}, \qquad
a'_3= a_{1}^{\dagger} + \sqrt{2}\; a_{3} ,
\end{equation}
with mode 3 denoting the idler mode. This amplifier is limited by the
quantum noise so it naturally leads to an optimal cloner.
A gain $G=2$ is needed since
the cloner doubles the energy by creating two clones 
with the same energy as the input state. 
One then produces these two clones simply 
by processing the output signal of the amplifier 
through a $50{\rm:}50$ phase-free beam splitter,
\begin{equation}  \label{imp2}
a'_{1}=\frac{1}{\sqrt{2}}(a_{out}+a_{2}), \qquad
a'_{2}=\frac{1}{\sqrt{2}}(a_{out}-a_{2}),
\end{equation}
as shown in Fig.~\ref{fig:ccircuit2}.
The rotation covariance of the resulting cloner is ensured by the  
fact that the amplifier and the beam splitter are phase-insensitive.
Actually, combining Eqs. (\ref{imp1}) and (\ref{imp2}) results in the same
canonical transformation as above, so this optical setup indeed implements
the optimal Gaussian cloner. It is readily checked that this setup 
leads to an equal $x$- and $p$-error variance of $1/2$ for both clones.

\section[]{Gaussian cloners with multiple inputs and outputs}

Let us now consider the general problem of optimal 
$N \to M$ cloning, extending what was done in \cite{gisi97} 
for the case of quantum bits. Consider a Gaussian transformation
which, from $N$ ($\geq 1$) identical replicas of an original input state, 
produces $M$ ($\geq 2$) output copies whose individual states 
are again given by an
expression similar to Eq. (\ref{eq:gaussmix}) but with an
error variance ${\sigma^2}_{N,M}$. (For the $1\to 2$ Gaussian cloner above,
we had ${\sigma^2}_{1,2}=1/2$.)  Using an argument based on the
concatenation of cloners, it is possible to derive a lower bound on
${\sigma^2}_{N,M}$, that is \cite{cerf00:coherent}
\begin{equation}
\label{eq:varclon}
\sigma^2_{N,M} \geq \frac{1}{N}-\frac{1}{M},
\end{equation}
so that the corresponding cloning fidelity for coherent states satisfies
\begin{equation}
\label{eq:fidclon}
F_{N,M} \leq \frac{MN}{MN+M-N}.
\end{equation}
The proof is connected to quantum state estimation theory,
the key idea being that cloning should not be a way of circumventing
the noise limitation encountered in any measuring process.
More specifically, concatenating a $N \to M$  cloner with
a $M \to L$ cloner results in a $N\to L$ cloner that cannot be better 
that the {\em optimal} $N \to L$ cloner. We then make use of the fact that
the excess noise variance of this $N\to L$ cloner simply is the sum
of the excess noise variances of the two component cloners
\cite{cerf00:coherent}. Denoting by $\sigma_{N,M}^2$ the excess noise
variance of the {\em optimal} $N\to M$ cloner, we get the inequality
$\sigma^2_{N,L} \leq \sigma^2_{N,M}+\sigma^2_{M,L}$.
In particular, if $L\to\infty$, we have
\begin{equation}\label{eq_addvar}
\sigma^2_{N,\infty}-\sigma^2_{M,\infty} \leq \sigma^2_{N,M}
\end{equation}
Since the limit of cloning with an infinite number of clones corresponds 
to a measurement, Eq.~(\ref{eq_addvar}) simply implies that cloning 
the $N$ replicas before measuring the $M$ resulting clones
does not provide a mean to enhance the accuracy of a direct measurement
of the $N$ replicas. This limit is useful
because the joint measurement of $x$ and $p$ on $N$ identical replicas
of a coherent state 
is known to give a minimum noise variance $\sigma^2_{N,\infty}=1/N$.
This, combined with Eq.~(\ref{eq_addvar}), 
gives the minimum noise variance induced by cloning, 
Eq.~(\ref{eq:varclon}), along with the corresponding cloning fidelity,
Eq.~(\ref{eq:fidclon}). Note that these bounds can
also be derived when $N=1$ using techniques similar to the ones used for
describing quantum nondemolition measurements. This was done in a
paper establishing a link between cloning and teleportation for
continuous variables~\cite{gros01}: for the $1\to 2$ cloner,
the teleportation fidelity must exceed $F_{1,2}=2/3$ in order to guarantee
that the teleported state is of better quality than the state kept by
the emitter.

\begin{figure}
\begin{center}
\includegraphics[width=3.0in,angle=0]{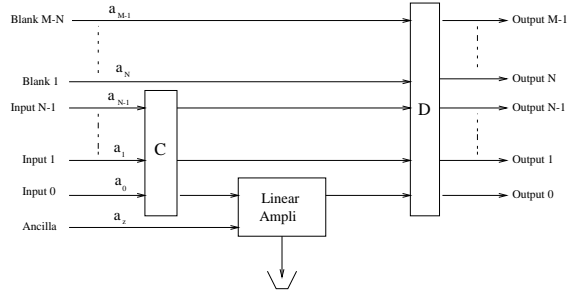}
\caption{Implementation of an $N \to M$ continuous-variable
cloning machine based on a phase-insensitive linear amplifier. 
Here, C stands for the amplitude concentration stage while
D refers to amplitude distribution. Both can be realized using a network
of beam-splitters that achieve a DFT. See \cite{brau00}.}
\label{fig:clonnm}
\end{center}
\end{figure}

Just like for the $1\to 2$ cloner, the bounds 
Eqs.~(\ref{eq:varclon}) and (\ref{eq:fidclon}) can be attained 
by a transformation whose implementation requires
only a phase-insensitive linear amplifier
and beam splitters \cite{brau00,fiur01}. Loosely speaking, the
procedure consists in concentrating the $N$ input modes into a single
mode by use of a network of beam splitters, then in amplifying the
resulting mode and distributing the output mode of the amplifier
into $M$ modes through a second network of beam-splitters. 
A convenient way to achieve these concentration and distribution stages
is provided by networks of beam splitters that realize
a Discrete Fourier Transform (DFT). Cloning is then
achieved by the following three-step procedure (see Fig.~\ref{fig:clonnm}).
First step: the $N$ input modes are concentrated into a single mode through a
DFT (acting on $N$ modes):
\begin{equation}
a'_k=\frac{1}{\sqrt{N}} \sum_{l=0}^{N-1} \exp(ikl 2\pi/N) \; a_l,
\end{equation}
with $k=0\ldots N-1$.
This operation concentrates the energy of the $N$ input modes $a_l$
into one single
mode $a_0'$ (hereafter renamed $a_0$) and leaves the remaining $N-1$ modes
($a'_1 \ldots a'_{N-1}$) in the vacuum state.
Second step: the mode $a_0$ is amplified with a linear amplifier
of gain $G=M/N$. This results in
\begin{eqnarray}
a'_0&=&\sqrt{\frac{M}{N}} \; a_0 + \sqrt{\frac{M}{N}-1} \; a_z^{\dagger},
\mathletter{a} \\
a'_z&=&\sqrt{\frac{M}{N}-1} \; a_{0}^{\dagger}+\sqrt{\frac{M}{N}} \; a_z.
\mathletter{b}
\end {eqnarray}
Third step: amplitude distribution by
performing a DFT (acting on $M$ modes) between the mode $a'_0$
and $M-1$ blank modes in the vacuum state:
\begin{equation}
a''_k=\frac{1}{\sqrt{M}} \sum_{l=0}^{M-1} \exp(ikl 2\pi/M) \; a'_l,
\end{equation}
with $k=0\ldots M-1$, and $a'_i=a_i$ for $i=N \ldots M-1$.
The DFT now distributes the energy
contained in the output of the amplifier among the $M$ output clones.  

It is readily checked that this procedure meets the requirements we put
on the $N\to M$ cloner, and is optimal. Indeed the 
quadrature variance of the $M$ output modes gives $1/2+1/N-1/M$, implying
that the cloning-induced excess noise variance is $1/N-1/M$. Furthermore,
the transformation is rotation covariant since the amplifier
and the beam splitters are phase insensitive. In conclusion,
we see that the optimal $N\to M$ cloning
transformation can be implemented using only passive elements
except for a single linear amplifier.

The above cloning transformation can be extended even further
by considering a generalized cloner that produces $M$ clones 
from $N$ replicas of a coherent state and $N'$ replicas
of its complex conjugate \cite{cerf01:pcic}. It is again
universal over the set of coherent states in the sense that
the cloning fidelities are invariant for all input coherent states. 
Interestingly, it can be shown that supplementing the
$N$ input states $\ket{\psi}^{\otimes N}$ with 
$N'$ phase-conjugated input states $\ket{\psi^*}^{\otimes N'}$
can, under certain circumstances, provide clones with a {\em higher} fidelity
than the above $N+N'\to M$ cloner. Note that, together with the $M$ clones,
this phase-conjugate input cloner also yields $M'$ anticlones 
(approximate copies of $\ket{\psi^*}$) at no cost, with $N-N'=M-M'$.
The advantage of having phase-conjugated inputs for a continuous-variable
cloner actually also has a counterpart in the context of qubit cloners. 
Indeed, motivated by this finding on continuous-variable cloners,
an optimal universal cloning transformation was recently 
derived that produces $M$ copies of an unknown pair of orthogonal 
qubits \cite{fiur02}.
For $M>6$, the cloning fidelity for a pair of orthogonal qubits can be shown
to be higher than that of the optimal cloning of a pair of identical qubits.
This is a first example of a quantum informational process that was initially
described for continuous-variable systems and only later on extended back
to quantum bits.

\section[]{Eavesdropping in continuous-variable quantum cryptography}

As mentioned above, quantum cloning can be viewed as an individual
eavesdropping strategy in continuous-variable quantum cryptography.
Consider a quantum key distribution scheme in which the key is
encoded into the displacement of a coherent or a squeezed state
that is drawn from a Gaussian distribution \cite{cerf00_qdgk,gros02_coherent}.
In the continuous-variable protocol defined in \cite{cerf00_qdgk}, 
which we will analyze here, squeezed states need to be used.
The emitter (Alice) prepares a squeezed state for which
the quadrature that is squeezed, $x$ or $p$, is chosen at random,
and then displaces it by $\hat D(r,0)$ or $\hat D(0,r)$ depending
on $x$ or $p$ is squeezed. Here, $r$ is drawn from a Gaussian
distribution, and constitutes a continuous key element. 
The receiver (Bob) then measures either
the $x$- or $p$-quadrature of the state he received, this choice
being again random. After Bob's measurement, Alice reveals the
quadrature she squeezed (and displaced) and Bob rejects the cases
where he measured the wrong quadrature, this discussion being made over
an authenticated public channel (this procedure is known as sifting).
The subset of states that are accepted by Bob then constitutes 
a Gaussian raw key (correlated Gaussian data at Alice's and Bob's side).
Indeed, denoting as $v$ the variance of the quadrature that is squeezed
by Alice, Bob gets for his measured quadrature an outcome $r'$ 
that is Gaussian distributed around $r$ with a variance $v$ (assuming
for the moment that the quantum channel is perfect and that 
there is no eavesdropping). 
If the variance of the random displacements $r$ imposed by Alice
is noted $V$, then this raw key shared by Alice and Bob
can be viewed as resulting from a Gaussian additive-noise channel
characterized by a signal-to-noise ratio of $V/v$.

The maximum amount of shared key bits that can be extracted
from this Gaussian raw key can be analyzed by applying
some standard notions of Shannon theory
for continuous channels [see {\it e.g.} \cite{cover}].
Consider a discrete-time continuous channel that adds a Gaussian noise
of variance $v$ to the signal. If the input $r$ of the channel 
is a Gaussian signal of variance $V$, 
the uncertainty on $r$ can be measured by its Shannon entropy
$h(r)=2^{-1} \log_2(2\pi\, {\rm e} \, V)$~bits.
Conditionally on $r$,
the output $r'$ is distributed as a Gaussian of variance $v$,
so that the entropy of $r'$ conditionally on $r$ becomes
$h(r'|r)= 2^{-1} \log_2(2\pi\, {\rm e} \, v)$~bits. Now,
the overall distribution of $r'$ is of course the convolution 
of these two distributions, {\it i.~e.}, a Gaussian of variance $V+v$, 
so that the output entropy is
$h(r') =2^{-1} \log_2(2\pi\, {\rm e} \, (V+v))$~bits.
According to Shannon theory, the information processed through
this noisy channel $r\to r'$ can be expressed as 
the amount by which the uncertainty on $r'$ is reduced by knowing $r$,
that is
\begin{equation}  \label{eq_shannon}
I\; {\rm (bits)}=h(r')-h(r'|r)= {1\over 2} \log_2\left(1+{V\over v}\right)
\end{equation}
where $V/v$ is the signal-to-noise ratio.
This is Shannon's famous formula for the capacity
of a Gaussian additive-noise channel. It is worth noticing
that this capacity is achieved in the case where the input 
is distributed as a Gaussian,
which is precisely the case under consideration here.

In the protocol analyzed in \cite{cerf00_qdgk}, the variances $v$ and $V$
are related by the constraint that Alice's choice of encoding the key
into either $x$ or $p$ should be invisible to a potential 
eavesdropper. In the first case, Alice applies a Gaussian-distributed 
displacement $\hat D(r,0)$ on a squeezed state whose $x$ quadrature has
a variance $v$, so that the quadratures $x$ and $p$ of this Gaussian mixture 
have a variance $V+v$ and $1/(4v)$, respectively. In the second case, 
Alice applies a displacement $\hat D(0,r)$ on a squeezed state in $p$,
resulting in a Gaussian mixture with variances $1/(4v)$ and $V+v$ 
for $x$ and $p$. These two Gaussian mixtures are required to be
indistinguishable, which simply translates into the requirement that
they have the same $x$ variances and the same $p$ variances:
\begin{equation}  \label{indistinguishable}
V+v={1\over 4v}
\end{equation}
This gives for the information
\begin{equation}
I = \log_2\left({1/2\over v}\right)
\end{equation}
which measures the maximum number of key bits that can be extracted
asymptotically (at the limit of long sequences) per use of the channel.
(The factor $1/2$ here is just the vacuum noise, so we see that
this protocol requires squeezing, that is, $v<1/2$.)
The actual methods that may be used to discretize the Gaussian raw
key and correct the resulting errors so as to extract a common bit string
are known as {\em reconciliation} protocols \cite{cerf02:epjd}.

Let us now consider the information that is transmitted in the
presence of an eavesdropper. We assume that the eavesdropper (Eve)
processes each key element into a Gaussian cloning machine,
keeps one clone, and sends the other one to Bob. Once the quadrature
that contains the key ($x$ or $p$) is revealed by Alice and Bob, Eve 
properly measures her clone. Clearly, Eve needs to use an asymmetric
version of the Gaussian cloner described above as she must be able
to tune the information she gains, and therefore the disturbance
she effects in the transmission. (A possible
implementation of this asymmetric Gaussian cloner is discussed 
in \cite{fiur01}.)
Thus, Eve adds some extra noise on the quadrature encoding the key, 
which results in a reduced signal-to-noise ratio on Alice-Bob channel.
Remember here, that the quality of the two clones
obey a no-cloning uncertainty relation akin to the Heisenberg
relation, implying that the product of the $x$-error variance on the
first clone times the $p$-error variance on the second one
remains bounded by $(1/2)^2$; 
see Eqs. (\ref{noclonineqa}) and (\ref{noclonineqb}). 
In particular, if $x$ and $p$ are treated symmetrically, we have
\begin{equation}  \label{eq-balance-noise}
\Delta n_B^2 \; \Delta n_E^2 \geq (1/2)^2
\end{equation}
This translates into a balance between the signal-to-noise ratio
in Alice-Bob channel $V/(v+\Delta n_B^2)$
and that in Alice-Eve channel $V/(v+\Delta n_E^2)$. This latter channel
is also a Gaussian channel so it can be treated similarly.
Using Eq.~(\ref{indistinguishable}), we can write the information
processed respectively in Alice-Bob and Alice-Eve channels as
\begin{eqnarray}
I_{AB} &=& {1\over 2} \log_2\left(
{ 1+4v\; \Delta n_B^2 \over 4v (v+\Delta n_B^2)} \right)   \mathletter{a} \\
I_{AE} &=& {1\over 2} \log_2\left(
{ 1+4v\; \Delta n_E^2 \over 4v (v+\Delta n_E^2)} \right)   \mathletter{b}
\end{eqnarray}
which gives
\begin{equation}
I_{AB}+I_{AE}-I={1\over 2} \log_2\left(
{(1+4v\; \Delta n_B^2)(1+4v\; \Delta n_E^2) \over 
4(v+\Delta n_B^2)(v+\Delta n_E^2)}
\right)
\end{equation}
One can then show that $I_{AB}+I_{AE}-I \leq 0$ by checking that the quantity
inside the logarithm is less or equal to one. This simplifies to the
condition
\begin{equation}
1-4v^2\leq 4\; \Delta n_B^2 \; \Delta n_E^2 \; (1-4v^2)
\end{equation}
which is indeed true as a consequence 
of Eq.~(\ref{eq-balance-noise}) and $v<1/2$.
Consequently, we have proven that, in this quantum cryptographic protocol, 
the no-cloning uncertainty relation translates 
into an information exclusion principle \cite{cerf00_qdgk}
\begin{equation}  \label{infoconservation}
I_{AB}+I_{AE} \leq I
\end{equation}
In other words, the information $I_{AE}$ gained by Eve 
is upper bounded by the defect of information at Bob's side, $I-I_{AB}$,
which implies that the security is guaranteed if $I_{AB}\geq I/2$
(since Bob then has an advantage over Eve, $I_{AB}\geq I_{AE}$).
Note that the bound in Eq.~(\ref{infoconservation}) is saturated 
by the asymmetric Gaussian cloner
discussed above, which strongly suggests that this is the optimal
individual attack (this actually can be proven rigorously). 
In practice, Alice and Bob can estimate the potentially eavesdropped 
information in the following way. 
Alice discloses the values $r$ she sent for a random
subset of the raw key. Then, Bob compares them to the values $r'$
he received, in order to estimate the variance of the distribution
of the differences $r'-r$, {\it i.~e.}, 
the excess noise variance $\Delta n_B^2$. This is sufficient to estimate
$I_{AB}$, and, via Eq.~(\ref{infoconservation}), an upper bound on $I_{AE}$.

An extended continuous-variable quantum key distribution 
protocol relying on Gaussian key carriers has recently been 
proposed in \cite{gros02_coherent}, where coherent states may be used 
instead of squeezed states. The encoding then consists in imposing 
a displacement $\hat D(x,p)$ onto the vacuum state
with $x$ and $p$ being drawn from a bi-variate Gaussian distribution.
Here, the choice of the quadrature is made by Bob, who decides 
to measure $x$ or $p$ at random, 
and then discloses his choice on the public channel.
The corresponding value of Alice's displacement ($x$ or $p$) together
with Bob's measured outcome again can be viewed as resulting from
a Gaussian channel, so the above information-theoretic treatment 
can be extended. In particular, one can calculate $I_{AB}$ and $I_{AE}$
in the case of an individual attack based on asymmetric Gaussian cloners.
The security analysis of this coherent-state protocol 
is beyond the scope of the present paper.

\begin{acknowledgments}
I would like to thank S.~L.~Braunstein, S.~Iblisdir, P.~van~Loock, 
S.~Massar, and G.~Van~Assche for their contribution
to the work reported on in this Chapter.
\end{acknowledgments}

\end{document}